\documentclass[11pt]{article}
\usepackage{amsmath}
\usepackage{amssymb}
\usepackage{cite}

\def\theequation{\thesection.\arabic{equation}}
\newtheorem{theorem}{Theorem}[section]

\title{\Large \textbf{Nonlocal symmetry of CMA generates ASD Ricci-flat metric with no Killing vectors}
\author{\large\textbf{M. B. Sheftel$^1$}\\[2mm]
 \normalsize $^1$~Department of Physics, Bo\u{g}azi\c{c}i University\\ \normalsize 34342 Bebek, Istanbul, Turkey
\footnote{E-mail: mikhail.sheftel@boun.edu.tr}}}

\date{}

\begin{document}
\maketitle

\begin{abstract}
The complex Monge-Amp\`ere equation $(CMA)$ in a two-component form is treated as a bi-Hamiltonian system.
 I present explicitly the first nonlocal symmetry flow in each of the two hierarchies of this system. An invariant solution of $CMA$ with respect to these nonlocal symmetries is constructed which, being a noninvariant solution in the usual sense, does not undergo symmetry reduction in the number of independent variables. I also construct the corresponding 4-dimensional anti-self-dual (ASD) Ricci-flat metric with either Euclidean or neutral signature. It admits no Killing vectors which is
one of characteristic features of the famous gravitational instanton $K3$. For the metric with the Euclidean signature, relevant for gravitational instantons, I explicitly calculate the Levi-Civita connection 1-forms and the Riemann curvature tensor.
\end{abstract}

\section{Introduction}

In his pioneer paper \cite{Plebanski}, Pleba\'nski demonstrated that anti-self-dual (ASD) Ricci-flat metrics on four-dimensional complex manifolds
are completely determined by a single scalar potential which satisfies his first or second heavenly equation. Such metrics are solutions to complex vacuum Einstein equations with zero cosmological constant. Real four-dimensional hyper-K\"ahler ASD metrics
\begin{equation}\label{metr}
  {\rm d} s^2 = u_{1\bar 1} {\rm d} z^1 {\rm d}\bar z^1 + u_{1\bar 2} {\rm d} z^1 {\rm d}\bar z^2 +
  u_{2\bar 1} {\rm d} z^2 {\rm d}\bar z^1 + u_{2\bar 2} {\rm d} z^2 {\rm d}\bar z^2
\end{equation}
that solve the vacuum Einstein equations with either Euclidean or
ultra-hyperbolic signature are governed by a scalar real-valued K\"ahler potential $u=u(z^1,z^2,\bar z^1,\bar z^2)$ which satisfies elliptic or hyperbolic complex Monge-Amp\`ere equation $(CMA)$
\begin{equation}
u_{1\bar 1}u_{2\bar 2} - u_{1\bar 2}u_{2\bar 1} = \varepsilon
 \label{cma}
\end{equation}
with $\varepsilon = \pm 1$ respectively. Here $u$ is a
real-valued function of the two complex variables $z^1 ,z^2$ and
their conjugates $\bar z^1 ,\bar z^2$, the subscripts denoting
partial derivatives with respect to these variables, e.g. $u_{1\bar 1} = \partial^2u/\partial z^1\partial\bar z^1$ and suchlike.
A modern proof of this result one can find in the books by Mason and Woodhouse \cite{Mason/Woodhouse} and Dunajski \cite{Dunajski}.

To illustrate this property, we introduce the coframe of one-forms
\begin{eqnarray}
&& \omega_1 = \frac{1}{\sqrt{u_{1\bar 1}}} (u_{1\bar 1}{\rm d}z_1 + u_{2\bar 1}{\rm d}z_2),\quad
\bar{\omega}_1  = \frac{1}{\sqrt{u_{1\bar 1}}} (u_{1\bar 1}{\rm d}\bar z_1 + u_{1\bar 2}{\rm d}\bar z_2)\nonumber\\
&& \omega_2 = \frac{1}{\sqrt{u_{1\bar 1}}} {\rm d}z_2,\quad \bar\omega_2 = \frac{1}{\sqrt{u_{1\bar 1}}} {\rm d}\bar z_2.
 \label{om_i}
\end{eqnarray}
The metric \eqref{metr} takes the canonical form
\begin{equation}
  {\rm d} s^2 =\omega_1\otimes\bar\omega_1 + \varepsilon\, \omega_2\otimes\bar\omega_2
 \label{omXom}
\end{equation}
where complex Monge-Amp\`ere equation \eqref{cma} has been used.
Equation \eqref{omXom} makes obvious the claim about the signature of the metric.

We are mostly interested in ASD Ricci-flat metrics that describe gravitational instantons which asymptotically look like a flat space, so that their curvature is concentrated in a finite region of a Riemannian space-time (see \cite{Dunajski,egh} and references therein). The most important gravitational instanton is $K3$ which geometrically is Kummer surface \cite{Atiyah/Hitchin/Singer}, for which an explicit form of the metric is still unknown while many of its properties and existence had been discovered and analyzed \cite{Hitchin1974,{Atiyah/Hitchin/Singer}, Yau,egh}.
A characteristic feature of the $K3$ instanton is that it does not admit any Killing vectors, that is, no continuous symmetries, which implies that the metric potential should be a noninvariant solution of $CMA$ equation.
As opposed to the case of invariant solutions, for noninvariant solutions of $CMA$ there should be no symmetry reduction \cite{olv} in the number of independent variables.
In this paper we achieve this goal by utilizing the invariance under nonlocal symmetries.

The main result of the paper has the form
\begin{eqnarray}
&& u = -\frac{\gamma}{c_0c_1^2}e^{-c_1x} + \frac{1}{3c_0^2c_1^2c_3^2\mu^2a^3} - \frac{c_2}{c_1}x + \frac{\gamma}{c_0c_1c_3}(y\sin\theta + z\cos\theta)\nonumber\\
&& + \frac{\gamma}{c_0c_1^2c_3^2}\left(\sigma_0 - \frac{\gamma\nu}{3c_0}\right) + \frac{\varepsilon\nu}{4c_1^2} + \rho_1t + \rho_2 + r(y,z),\qquad v = u_t
 \label{main}
\end{eqnarray}
where
\begin{equation}
  a = \mp \Gamma^{-1/2}, \quad \Gamma = 2c_0\mu\left(c_3^2e^{-c_1x} - \sigma(y,z,t)\right) + \nu^2
 \label{aa}
\end{equation}
with an arbitrary harmonic function $r(y,z)$ and  $\sigma$, $\sigma_0$ defined in \eqref{sigm}, \eqref{si0}, respectively, $\gamma = \nu/\mu$, $\mu$ and $\nu$ defined by \eqref{mu} and \eqref{nu}, $\theta = -c_1t + \theta_0$, $c_0,c_1,c_2,c_3,\rho_1,\rho_2,\theta_0$ are arbitrary real constants.
\begin{theorem}
The expressions \eqref{main}, \eqref{aa} for $u$ and $v$ provide a solution of the $CMA$ in its real two-component form \eqref{2comp}. This solution by construction is
invariant under the first nonlocal symmetry flow of the $CMA$ in each of the two hierarchies of this equation. With the generic set of constants, this solution is noninvariant under point symmetries of $CMA$ and consequently it experiences no reduction in the number of independent variables.
\end{theorem}
Noninvariance of the solution for the K\"ahler potential of the metric implies the absence of point symmetries (Killing vectors) for the metric itself since no reduction in the number of independent variables obviously takes place also for the metric coefficients in Section \ref{sec-metr}.

The noninvariance of the solution \eqref{main} under  point symmetries looks obvious from the absence  of symmetry reduction in the solution which clearly depends on four independent combinations of variables $t,x,y,z$ (two different combinations of $y$ and $z$ enter expressions for $v = \partial u/\partial t$ and there is obviously no reduction in either $x$ or $t$). The outline of a more rigorous approach to the proof of noninvariance is formulated in the Appendix.

As far as we know, there are no published examples of noninvariant solutions of $CMA$ and the corresponding gravitational metrics with no Killing vectors
with the exception of our earlier paper \cite{S-M}, and there was still no progress in the explicit construction of $K3$ instanton metric. Instructive reviews of the present state of the theory of gravitational instantons can be found in the book \cite{Dunajski} and review \cite{egh}.

The paper is organized as follows. In Section \ref{sec-real}, we convert $CMA$ equation into real variables and two-component form. In Section \ref{sec-recursion}, we exhibit two alternative bi-Hamiltonian representations of $CMA$ which we discovered earlier \cite{nsky}. In Section \ref{sec-high},
we explicitly construct first nonlocal flows in each hierarchy of $CMA$ system related to its two bi-Hamiltonian structures.

In Section \ref{sec-invar},
we formulate the invariance conditions with respect to both nonlocal symmetries appearing in each of the two alternative hierarchies. In this way, we keep in the invariance conditions the obvious discrete symmetry relating two bi-Hamiltonian structures. Thus, alternatively  we can say that we impose invariance conditions under one nonlocal symmetry and the discrete symmetry. In the following, for simplicity we restrict ourselves to the invariance under the \textit{special first nonlocal symmetry} by setting $\Phi=0$ and $\chi=0$.

In Section \ref{sec-intcond}, we show in detail how a careful analysis of integrability conditions specifies various functional parameters in the invariance equations with no additional assumptions made. In Section \ref{sec-integr}, we integrate completely all the obtained equations and end up with a noninvariant solution of $CMA$ which is a general form of the solution invariant under the special first nonlocal symmetry in each hierarchy. In Section \ref{sec-metr}, we use this solution for constructing the corresponding (anti-)self-dual Ricci-flat metric with either Euclidean or neutral signature.

In Section \ref{sec-connect} we calculate Levi-Civita connection 1-forms for Euclidean signature by solving Cartan structure equations in the case of no torsion. In Section \ref{sec-curv} we obtain the corresponding Riemann curvature tensor components.
These results should be helpful for the future analysis of singularities and asymptotic behavior of the metric.
We note that the results for the connection and Riemann curvature are not only related to our solution but are also valid more generally for the K\"ahler metric \eqref{metrQ} generated by any solution of $CMA$ which may be helpful for future researchers dealing with this subject.
In the Appendix, we give the full set of point symmetries of $CMA$ and outline the approach needed for the rigorous proof of noninvariance of its solutions.

\section{Real variables and 2-component form of CMA}
\setcounter{equation}{0}
 \label{sec-real}

In our earlier paper \cite{nsky} we presented bi-Hamiltonian
structure of the two-component version of \eqref{cma}, which
by Magri's theorem \cite{magri} proves that it is a
completely integrable system in four dimensions.

We impose additional reality condition for all the objects in the theory. The transformation from complex
to real variables has the form
\begin{equation}
 t = z^1 + \bar z^1,\quad x = i(\bar z^1 - z^1),\quad y = z^2 + \bar z^2,\quad z = i(\bar z^2 - z^2).
\label{realvar}
\end{equation}
We introduce the notation
\begin{equation}
 a = \Delta(u) = u_{yy} + u_{zz},\quad b = u_{xy} - v_z,\quad c = v_y +   u_{xz},\quad Q = \frac{b^2 + c^2 + \varepsilon}{a}
 \label{not}
\end{equation}
where $v = u_t$ is the second component of the unknown and $\Delta = D_y^2 + D_z^2$ is the two-dimensional
Laplace operator (letter subscripts denote corresponding partial derivatives). The definitions \eqref{not} imply the relations
\begin{equation}
 a_x = b_y + c_z,\quad c_y - b_z = \Delta(v) \equiv a_t.
 \label{rel}
\end{equation}

The $CMA$ equation \eqref{cma} in the real variables becomes
\begin{equation*}
 (u_{tt} + u_{xx})\Delta(u) - b^2 - c^2 - \varepsilon = 0
\end{equation*}
or in the two-component form
\begin{equation}
 \left(\begin{array}{c}
  u_t\\ v_t
 \end{array}\right) =
 \left(\begin{array}{c}
  v\\ Q - u_{xx}
 \end{array}\right)
 \label{2comp}
\end{equation}
which we will call $CMA$ system.

The metric \eqref{metr} in real variables reads
\begin{eqnarray}
&& {\rm d} s^2 = (v_t + u_{xx})({\rm d} t^2 + {\rm d} x^2) + a({\rm d} y^2 + {\rm d} z^2) \nonumber\\
&&\mbox{} - 2b({\rm d} t\,{\rm d} z  - {\rm d} x\,{\rm d} y) + 2c({\rm d} t\,{\rm d} y  + {\rm d} x\,{\rm d} z).
 \label{realmetr}
\end{eqnarray}
The coframe of one-forms becomes
\begin{eqnarray}
&& \Omega_1 = \frac{1}{\sqrt{v_t + u_{xx}}}[(c + ib)({\rm d}y + i{\rm d}z) + (v_t + u_{xx})({\rm d}t + i{\rm d}x)]\nonumber\\
&& \Omega_2 = \frac{{\rm d}y + i{\rm d}z}{\sqrt{v_t + u_{xx}}}
 \label{Om}
\end{eqnarray}
with the metric
\begin{equation}
  {\rm d} s^2 =\Omega_1\otimes\bar\Omega_1 + \varepsilon\, \Omega_2\otimes\bar\Omega_2.
 \label{OmXOm}
\end{equation}

\section{Bi-Hamiltonian representations of CMA system}
\setcounter{equation}{0}
 \label{sec-recursion}

The $CMA$ system \eqref{2comp} can be put in the Hamiltonian form
\begin{equation}
\left(
\begin{array}{c}
u_t \\ v_t
\end{array}
\right) =  J_0 \left(
\begin{array}{c}
\delta_u H_1 \\ \delta_v H_1
\end{array}
\right)
 \label{Hamilton}
\end{equation}
where $\delta_u$ and $\delta_v$ are Euler-Lagrange operators
\cite{olv} with respect to $u$ and $v$. Here $J_0$ is the Hamiltonian operator
\begin{equation}
 J_0 =  \left(
\begin{array}{cc}
\phantom{-} 0 & \hspace*{2mm}\frac{\textstyle 1}{\textstyle a}
\\[2mm] -\frac{\textstyle 1}{\textstyle a} & \hspace*{2mm}\frac{\textstyle 1}{\textstyle a}
(cD_y + D_yc - bD_z - D_z b)\frac{\textstyle 1}{\textstyle a}
\end{array}
\right) \label{J0R}
\end{equation}
determining the structure of Poisson bracket and $H_1$ is the corresponding Hamiltonian density
\begin{equation}\label{H1R}
  H_1 = \frac{1}{2}\,[v^2\Delta(u) - u_{xx}(u_y^2 + u_z^2)] -
  \varepsilon u.
\end{equation}

The first real recursion operator has the form
\begin{eqnarray}\label{R1}
 & & R_1 = \left(
  \begin{array}{cr}
  0 & 0
  \\ QD_z - cD_x & b
  \end{array}
  \right) +
  \\ & &
 \Delta^{-1}\!\circ\left(
  \begin{array}{cc}
 D_y \Bigl(- a D_x + bD_y + cD_z \Bigr) + D_z\Bigl(cD_y -
 bD_z\Bigr)
 & - D_z a
 \\[2mm] \begin{array}{l}
  D_x \Bigl[D_y\Bigl(cD_y - bD_z\Bigr) + D_z\Bigl(aD_x - bD_y -
 cD_z\Bigr)\Bigr]
 \end{array}
 & - D_xD_y a
  \end{array}
  \right) \nonumber
\end{eqnarray}
where $\circ$ means operator multiplication, and the second
recursion operator reads
\begin{eqnarray}\label{R2}
 & & R_2 = \left(
  \begin{array}{cr}
   0 & 0 \\
   bD_x - QD_y & c
  \end{array}
  \right) +
  \\ & & \Delta^{-1}\circ
  \left(
  \begin{array}{lr}
 D_y(bD_z - cD_y) + D_z(-aD_x + bD_y + cD_z)
 & D_y a
  \\ D_x\Bigl[D_y(- aD_x + bD_y + cD_z) + D_z(cD_y - bD_z)\Bigr]
 & - D_xD_z a
  \end{array}
  \right). \nonumber
\end{eqnarray}
The two recursion operators $R_1$ and $R_2$ generate two alternative second Hamiltonian operators $J_1 = R_1J_0$ and $J^1 = R_2J_0$
 \begin{eqnarray}\label{J1}
& & J_1 = R_1J_0 = \Delta^{-1}\circ\left(
\begin{array}{cc}
 D_z & -D_xD_y
\\ D_xD_y & D_x^2D_z
\end{array}
 \right) +
 \\[2mm] & & \left(\begin{array}{cc}
 \phantom{-} 0 & \frac{\textstyle b}{\textstyle a}
 \\[2mm] -\frac{\textstyle b}{\textstyle a} &
 \frac{\textstyle c}{\textstyle a^2}\Bigl(bD_y - aD_x\Bigr)
 + \Bigl(D_y b - D_x a\Bigr)\frac{\textstyle c}{\textstyle a^2}
 + \frac{\textstyle Q_-}{\textstyle 2a}D_z
 + D_z\frac{\textstyle Q_-}{\textstyle 2a}
\end{array}\right) \nonumber
 \end{eqnarray}
where $Q_- = (c^2 - b^2 + \varepsilon)/a$, and
 \begin{eqnarray}\label{J^1}
& & J^1 = R_2J_0 = \Delta^{-1}\circ\left(
\begin{array}{cc}
 D_y & D_xD_z
\\ - D_xD_z & D_x^2D_y
\end{array}
 \right) +
 \\[2mm] & & \left(\begin{array}{cc}
 \phantom{-} 0 & - \frac{\textstyle c}{\textstyle a}
 \\[2mm] \frac{\textstyle c}{\textstyle a} &
 \frac{\textstyle b}{\textstyle a^2}\Bigl(cD_z - aD_x\Bigr)
 + \Bigl(D_z c - D_x a\Bigr)\frac{\textstyle b}{\textstyle a^2}
 + \frac{\textstyle Q^-}{\textstyle 2a}D_y
 + D_y\frac{\textstyle Q^-}{\textstyle 2a}
\end{array}\right) \nonumber
 \end{eqnarray}
where $Q^- = (b^2 - c^2 + \varepsilon)/a$.

The flow (\ref{Hamilton}) can be generated by the Hamiltonian operator
$J_1$ from the Hamiltonian density
\begin{equation}\label{H0}
  H_0 = zv\Delta(u) + u_xu_y
\end{equation}
so that CMA in the two-component form (\ref{Hamilton}) is a {\it bi-Hamiltonian system} \cite{magri}
\begin{equation}\label{biHam}
\left(
\begin{array}{c}
u_t \\ v_t
\end{array}
\right) =  J_0 \left(
\begin{array}{c}
\delta_u H_1 \\ \delta_v H_1
\end{array}
\right) =  J_1 \left(
\begin{array}{c}
\delta_u H_0 \\ \delta_v H_0
\end{array}
\right).
\end{equation}

The same flow (\ref{Hamilton}) can also be generated by the
Hamiltonian operator $J^1$ from the Hamiltonian density
\begin{equation}\label{H^0}
  H^0 = yv\Delta(u) - u_xu_z
\end{equation}
which yields another bi-Hamiltonian representation of the $CMA$
system (\ref{Hamilton})
\begin{equation}\label{biHam2}
\left(
\begin{array}{c}
u_t \\ v_t
\end{array}
\right) =  J_0 \left(
\begin{array}{c}
\delta_u H_1 \\ \delta_v H_1
\end{array}
\right) =  J^1 \left(
\begin{array}{c}
\delta_u H^0 \\ \delta_v H^0
\end{array}
\right).
\end{equation}
The relation between the two bi-Hamiltonian structures is realized by the \textit{discrete symmetry}  $z\mapsto y, y\mapsto -z$.

\section{Nonlocal flows}
\setcounter{equation}{0}
 \label{sec-high}

The first nonlocal flows of each hierarchy of $CMA$ system are generated by $J_1$ and $J^1$
acting on the vector of variational derivatives of $H_1$
\begin{equation}\label{flow1}
 \left(
\begin{array}{c}
u_{\tau_2} \\ v_{\tau_2}
\end{array}
\right) =  J_1 \left(
\begin{array}{c}
\delta_u H_1 \\ \delta_v H_1
\end{array}
\right)
\end{equation}
\begin{equation}\label{flow1'}
 \left(
\begin{array}{c}
u_{\tau_{2^{\,\prime}}} \\ v_{\tau_{2^{\,\prime}}}
\end{array}
\right) =  J^1 \left(
\begin{array}{c}
\delta_u H_1 \\ \delta_v H_1
\end{array}
\right)
\end{equation}
where $\tau_2$, $\tau_{2'}$ are time variables of the flows \eqref{flow1}, \eqref{flow1'}, respectively.
Using the expressions \eqref{J1}, \eqref{J^1} and \eqref{H1R} for $J_1$, $J^1$ and $H_1$ we obtain explicit expressions
for the flows \eqref{flow1} and \eqref{flow1'}
\begin{eqnarray}
 && u_{\tau_2} = \Delta^{-1}\left\{D_z(au_{xx} - u_{xy}^2 - u_{xz}^2 - \varepsilon) - D_xD_y(a v)\right\} + u_{xy}v\nonumber\\
 && u_{\tau_{2^{\,\prime}}} = \Delta^{-1}\left\{D_y(au_{xx} - u_{xy}^2 - u_{xz}^2 - \varepsilon) + D_xD_z(a v)\right\} - u_{xz}v.
 \label{nonloc}
\end{eqnarray}
Second components of these flows are time derivatives of \eqref{nonloc}, $v_{\tau_2} = D_t[u_{\tau_2}]$, $v_{\tau_{2^{\,\prime}}} = D_t[u_{\tau_{2^{\,\prime}}}]$, so that the flows \eqref{flow1} and \eqref{flow1'} commute with the flow \eqref{2comp} of $CMA$ system
and hence they are nonlocal symmetries of the $CMA$ system.

\section{Invariance conditions with respect to nonlocal symmetries}
\setcounter{equation}{0}
 \label{sec-invar}

Solutions invariant with respect to nonlocal symmetries are determined by the conditions $u_{\tau_2} = 0$ and $u_{\tau_{2^{\,\prime}}} = 0$ which due to \eqref{nonloc} take the explicit form (with $a$ replaced by  $\Delta[u]$ according to \eqref{not})
\begin{eqnarray}
 && D_z\left[\Delta[u]u_{xx} - u_{xy}^2 - u_{xz}^2\right] - D_xD_y\left[v\Delta[u]\right] + \Delta\left[vu_{xy}\right] = 0 \nonumber\\
 && D_y\left[\Delta[u]u_{xx} - u_{xy}^2 - u_{xz}^2\right] + D_xD_z\left[v\Delta[u]\right] - \Delta\left[vu_{xz}\right] = 0.
 \label{invar}
\end{eqnarray}
 Here we impose both invariance conditions \eqref{invar} on solutions of the $CMA$ system in order to keep the \textit{discrete symmetry}  $z\mapsto y, y\mapsto -z$ between the two bi-Hamiltonian structures. Differentiating the first and second equations \eqref{invar} with respect to $y$ and $z$, respectively, and taking the difference of the results yield the integrability condition
\begin{equation*}
 \Delta\left\{D_y[vu_{xy}] + D_z [vu_{xz}] - D_x[v\Delta[u]]\right\} = 0
\end{equation*}
or, equivalently
\begin{eqnarray}
&& \hspace*{-4.5pt} D_y[vu_{xy}] + D_z [vu_{xz}] - D_x[v\Delta[u]] = \Phi(x,y,z,t) \nonumber\\
&& \iff v_yu_{xy} + v_zu_{xz} - v_x\Delta[u] = \Phi,\quad \textrm{where}\quad \Delta[\Phi] = 0.\nonumber\\
\label{integr}
\end{eqnarray}
On account of the relation \eqref{integr} each of the relations \eqref{invar} becomes
\begin{equation}
 v_yu_{xz} - v_zu_{xy} = \Delta[u]u_{xx} - u_{xy}^2 - u_{xz}^2 + \chi(x,y,z,t)
 \label{invari}
\end{equation}
where $\Phi_y = \chi_z$ and $\Phi_z = - \chi_y$ and hence $\Delta[\Phi]=\Delta[\chi]=0$.
Thus, we end up with the system of two equations \eqref{integr} and \eqref{invari}   linear in derivatives of $v$. Solving this system algebraically for $v_y$ and $v_z$ and denoting $\delta = u_{xy}^2 + u_{xz}^2$, we obtain
\begin{eqnarray}
&& v_y = \frac{1}{\delta}\left\{\Delta[u](v_xu_{xy} + u_{xx}u_{xz}) - \delta u_{xz} + \Phi u_{xy} + \chi u_{xz}\right\}\nonumber\\
&& v_z = \frac{1}{\delta}\left\{\Delta[u](v_xu_{xz} - u_{xx}u_{xy}) + \delta u_{xy} + \Phi u_{xz} - \chi u_{xy}\right\}.
 \label{linsol}
\end{eqnarray}
In the following for simplicity we set $\Phi=0$, $\chi=0$ and refer to this case as the invariance under \textit{special first nonlocal symmetries}. In the following it is convenient to introduce the quantity
\begin{equation}
 w = \frac{\delta}{\Delta[u]} - u_{xx}.
\label{w}
\end{equation}
Equations \eqref{linsol} become
\begin{equation}
  v_y = \frac{u_{xy}v_x-u_{xz}w}{w+u_{xx}},\quad v_z = \frac{u_{xz}v_x+u_{xy}w}{w+u_{xx}}
  \label{linsimpl}
\end{equation}
with the immediate consequences
\begin{equation}
  v_x = \frac{1}{\Delta[u]}\,  (u_{xy}v_y + u_{xz}v_z),\quad w = \frac{1}{\Delta[u]}\, (u_{xy}v_z - u_{xz}v_y).
 \label{v_x(v_y,v_z)}
\end{equation}

On account of the equations \eqref{w} and \eqref{linsimpl}, the real two-component form \eqref{2comp} of $CMA$ becomes
\begin{equation}
 v_t = \frac{v_x^2 - u_{xx}w}{w+u_{xx}} + \frac{\varepsilon}{\Delta[u]}.
 \label{cmareal}
\end{equation}

Integrability condition $(v_y)_z - (v_z)_y = 0$ of equations \eqref{linsimpl} yields
\begin{equation}
  (u_{xz}w_y - u_{xy}w_z) v_x = u_{xx}(u_{xy}w_y + u_{xz}w_z) - \Delta[u](w+u_{xx})w_x.
 \label{v_x}
\end{equation}

\section{Further integrability conditions}
\setcounter{equation}{0}
 \label{sec-intcond}

It is convenient to take equation \eqref{v_x} in the form
\begin{equation}
  (u_{xz}v_x - u_{xx}u_{xy})w_y = (u_{xy}v_x + u_{xx}u_{xz})w_z - \Delta[u](w + u_{xx})w_x.
 \label{wyzx}
\end{equation}
The integrability conditions $(v_t)_y = (v_y)_t$ and $(v_t)_z = (v_z)_t$ with the use of \eqref{wyzx} simplify to
\begin{equation}
  \Delta[u_y] = \frac{u_{xy}}{w+u_{xx}} \Delta[u_x],\quad \Delta[u_z] = \frac{u_{xz}}{w+u_{xx}} \Delta[u_x]
 \label{v_tyz}
\end{equation}
with the integrability condition $(\Delta[u_y])_z - (\Delta[u_z])_y = 0$ resulting in
\begin{equation}
  u_{xz}w_y - u_{xy}w_z = 0.
 \label{w_yz}
\end{equation}
On account of \eqref{w_yz} equation \eqref{v_x} becomes
\begin{equation}
  u_{xx}(u_{xy}w_y + u_{xz}w_z) - \Delta[u](w+u_{xx})w_x = 0.
 \label{w_xyz}
\end{equation}
With the aid of the definition \eqref{w}, equations \eqref{w_yz} and \eqref{w_xyz} can be put in the form
\begin{equation}
  w_y = \frac{u_{xy}}{u_{xx}} w_x,\quad w_z = \frac{u_{xz}}{u_{xx}} w_x
 \label{wy_wz}
\end{equation}
with the integrability condition $(w_y)_z=(w_z)_y$ identically satisfied.
Differentiating the definition \eqref{w} of $w$ with respect to $t$ we obtain
\begin{equation}
  w_t =\frac{v_x}{u_{xx}} w_x.
 \label{w_t}
\end{equation}
The integrability conditions $(w_t)_y=(w_y)_t$ and $(w_t)_z=(w_z)_t$ of equations \eqref{wy_wz} and \eqref{w_t} are identically satisfied.

Equations \eqref{wy_wz} are integrated by the method of characteristics with the result $w=w(u_x,t)$ whereas  \eqref{w_t} further implies $w=w(u_x)$. The remaining equations \eqref{v_tyz}
take the form of two first-order PDEs for $\Delta[u]$
\begin{eqnarray}
&&  (u_{xx} + w(u_x))(\Delta[u])_y - u_{xy}(\Delta[u])_x = 0\nonumber\\
&&  (u_{xx} + w(u_x))(\Delta[u])_z - u_{xz}(\Delta[u])_x = 0.
 \label{vt_yz}
\end{eqnarray}
We start with the combination of these equations
\begin{equation}
  u_{xz}(\Delta[u])_y - u_{xy}(\Delta[u])_z = 0
 \label{vyz}
\end{equation}
which is integrated by the method of characteristics to give\newline $\Delta[u]=f(u_x,x,t)$. Then equations \eqref{vt_yz} are satisfied by the solution
\begin{equation}
  \Delta[u] = f(\zeta,t), \quad \textrm{where}\quad \zeta = \omega(u_x) + x \quad \textrm{and}\quad \omega(u_x) = \int\frac{d\,u_x}{w(u_x)}.
 \label{Delu}
\end{equation}
Equation \eqref{w} becomes
\begin{equation}
 \delta \equiv u_{xy}^2 + u_{xz}^2 = A^2,\quad \textrm{where}\quad A = \sqrt{f\left(u_{xx} + \frac{1}{\omega'}\right)}.
 \label{A}
\end{equation}
We rewrite \eqref{A} in the form
\begin{equation}
  u_{xy} = A\sin\theta, \quad u_{xz} = A\cos\theta
 \label{th}
\end{equation}
where we have introduced the new unknown $\theta=\theta(x,y,z,t)$.
The definition of $A$ in \eqref{A} implies
\begin{eqnarray}
&& A_y = \tilde{a}\sin\theta + \frac{f}{2} \cos\theta\cdot\theta_x, \quad A_z = \tilde{a}\cos\theta - \frac{f}{2} \sin\theta\cdot\theta_x \nonumber\\
&& A_x = \frac{f_\zeta A^3}{2f^2}\omega' - \frac{A}{2}\frac{\omega''}{\omega'^2} + \frac{f}{2A}\left(\frac{\omega''}{\omega'^3} + u_{xxx}\right)
 \label{A_}
\end{eqnarray}
where
\begin{equation}
  \tilde{a} = \frac{3}{4}\left(\frac{f_\zeta}{f} A^2\omega' - f\frac{\omega''}{\omega'^2}\right) + \frac{f^2}{4A^2}\left(\frac{\omega''}{\omega'^3} + u_{xxx}\right).
 \label{a}
\end{equation}
The integrability condition $(u_{xy})_z - (u_{xz})_y = 0$ of the system \eqref{th} has the form
\begin{equation}
  \sin\theta\cdot \theta_y + \cos\theta\cdot \theta_z = \frac{f}{2A} \theta_x.
 \label{int_th}
\end{equation}

The first equation $\Delta[u] = f(\zeta,t)$ in \eqref{Delu} implies
\begin{equation}
  \Delta[u_x] = (u_{xy})_y + (u_{xz})_z = (A\sin\theta)_y + (A\cos\theta)_z = f_\zeta(\omega'u_{xx} + 1)
 \label{Delu_x}
\end{equation}
which finally results in the equation
\begin{equation}
  \cos\theta\cdot \theta_y - \sin\theta\cdot \theta_z = \frac{f_\zeta A}{f}\omega' - \frac{\tilde{a}}{A}.
 \label{eqth2}
\end{equation}

We solve algebraically the system of equations \eqref{int_th} and \eqref{eqth2} to obtain
\begin{equation}
  \theta_y = \frac{f}{2A} \sin\theta\cdot\theta_x + \tilde{b}\cos\theta,\quad \theta_z = \frac{f}{2A} \cos\theta\cdot\theta_x - \tilde{b}\sin\theta
 \label{th_yz}
\end{equation}
where
\begin{equation}
  \tilde{b} = \frac{f_\zeta}{f} A\omega' - \frac{\tilde{a}}{A}.
 \label{b}
\end{equation}
Integrability condition for the system \eqref{th_yz} reads
\begin{equation}
  (\theta_y)_z - (\theta_z)_y = \tilde{b}_z\cos\theta + \tilde{b}_y\sin\theta - \frac{f}{2A} \tilde{b}_x + \tilde{b}^2 + \frac{f^2}{4A^2} \theta_x^2 = 0
 \label{thyz}
\end{equation}
or in the explicit form
\begin{eqnarray}
&& \left(\frac{f^2}{2A^3}u_{xxx} - \frac{f_\zeta}{2f}A\omega' + \frac{f^2}{2A^3}\frac{\omega''}{\omega'^3} - \frac{3f}{2A}\frac{\omega''}{\omega'^2}\right)^2
+ \frac{f^2}{A^2} \theta_x^2 = 0.
\label{compcond}
\end{eqnarray}
Reality condition for \eqref{compcond} implies that both quadratic terms vanish separately
\begin{equation}
  u_{xxx} = \frac{f_\zeta}{f^3}A^4\omega' - \frac{\omega''}{\omega'^3} + \frac{3A^2}{f}\frac{\omega''}{\omega'^2}
 \label{compat}
\end{equation}
and $\theta_x = 0\iff \theta=\theta(y,z,t)$. Using \eqref{compat} in \eqref{a} for $\tilde{a}$  and then in \eqref{b} for $\tilde{b}$ we obtain
\begin{equation}
 \tilde{b} = 0,\quad \tilde{a}  = \frac{f_\zeta}{f} A^2\omega'
 \label{b0a}
\end{equation}
and \eqref{th_yz} with $\theta_x=0$ implies $\theta_y = \theta_z = 0$, so that $\theta = \theta(t)$. Equations \eqref{A_} become
\begin{equation}
  A_y = \tilde{a}\sin\theta,\quad A_z = \tilde{a}\cos\theta,\quad A_x = \frac{f_\zeta}{f^2} A^3\omega' + A\frac{\omega''}{\omega'^2}.
 \label{der_A}
\end{equation}
From the definition \eqref{A} of $A$ it follows
\begin{equation}
  u_{xx} = \frac{A^2}{f} - \frac{1}{\omega'}
 \label{uxx}
\end{equation}
Since $f=f(\zeta,t)$ where $\zeta = \omega(u_x) + x$, we have
\begin{equation}
   f_x = \frac{f_\zeta}{f} A^2 \omega',\quad f_y = f_\zeta \omega'A\sin\theta \quad f_z = f_\zeta \omega'A\cos\theta
 \label{der_f}
\end{equation}
and hence $A_y/A = f_y/f$, $A_z/A = f_z/f$, so that $A = \alpha(x,t)f$.
Then the last equation in \eqref{der_A} implies
\begin{equation}
  \frac{A_x}{A} = \frac{f_x}{f} + \frac{\omega''}{\omega'^2} = \frac{f_x}{f} +\frac{\alpha_x(x,t)}{\alpha}
  \Longrightarrow\left(\frac{\omega''}{\omega'^2}\right)_y=0, \quad\left(\frac{\omega''}{\omega'^2}\right)_z=0
 \label{der_om}
\end{equation}
and hence
\begin{equation}
  \left(\frac{\omega''}{\omega'^2}\right)' = 0 \Longrightarrow \frac{1}{\omega'} = c_1u_x + c_2 \iff \omega' = \frac{1}{c_1u_x + c_2}.
 \label{om}
\end{equation}
From \eqref{der_om} and \eqref{om} it follows that
\begin{equation}
  \alpha(x,t) = c_3(t) e^{-c_1(t)x},\quad A = c_3(t)e^{-c_1 x}f(\zeta,t)
 \label{al}
\end{equation}
while \eqref{uxx} and \eqref{th} imply
\begin{eqnarray}
&& u_{xx} = c_3^2(t)e^{-2c_1 x} f - (c_1u_x + c_2) \nonumber\\
&& u_{xy} = c_3e^{-c_1x}f\sin\theta,\quad u_{xz} = c_3e^{-c_1x}f\cos\theta
 \label{u_xx}
\end{eqnarray}
while  \eqref{compat} takes the form
\begin{equation}
  u_{xxx} = c_3^4(t)\frac{e^{-4c_1x} ff_\zeta}{c_1u_x+c_2} - 3c_3^2(t)c_1e^{-2c_1x} f + c_1(c_1u_x+c_2).
 \label{u_xxx}
\end{equation}
It is easy to check that the integrability condition $(u_{xx})_x = u_{xxx}$ of equations \eqref{u_xx} and \eqref{u_xxx} is identically satisfied.

Expressions \eqref{linsimpl} and \eqref{cmareal} for derivatives of $v$ take the form
\begin{eqnarray}
&& v_y = \frac{e^{c_1x}}{c_3(t)}[v_x\sin\theta - (c_1u_x+c_2)\cos\theta]\nonumber\\[2pt]
&& v_z = \frac{e^{c_1x}}{c_3(t)}[v_x\cos\theta + (c_1u_x+c_2)\sin\theta]\nonumber\\[2pt]
&& v_t = \frac{e^{2c_1x}}{c_3^2(t)f}[v_x^2 + (c_1u_x+c_2)^2] - (c_1u_x+c_2) + \frac{\varepsilon}{f}.
 \label{v_}
\end{eqnarray}
Integrability condition $(v_t)_x = (v_x)_t$ where $v_x$ is determined by \eqref{v_x(v_y,v_z)} in the form
\[v_x = c_3(t)e^{-c_1x}(v_y\sin\theta + v_z\cos\theta)\]
 yields the relation
\begin{equation}
  (c_1 + \theta'(t))(c_1u_x+c_2) - \frac{c_3'(t)}{c_3(t)} v_x = 0.
 \label{vtx}
\end{equation}
If $c_3'(t)\ne0$, \eqref{vtx} can be solved for $v_x$ and the integrability conditions $(v_y)_x = (v_x)_y$, $(v_z)_x = (v_x)_z$ imply $c_1c_3=0$ and, since $c_1\ne0$ (otherwise \eqref{vtx} leads to a reduction), we have $c_3=0$ which contradicts to our assumption $c_3'(t)\ne0$.

Thus, we proceed to the opposite case of constant $c_3$ in \eqref{vtx} which implies $\theta(t) = -c_1 t + \theta_0$ with $v_x$ remaining undetermined. Then we check that all the integrability conditions $(v_y)_t = (v_t)_y$,  $(v_z)_t = (v_t)_z$ and $(v_y)_z = (v_z)_y$ are identically satisfied.

Differentiating first two equations \eqref{v_} with respect to $x$ we obtain
\begin{eqnarray}
&& v_{xy} = \frac{e^{c_1x}}{c_3}(v_{xx} + c_1v_x)\sin\theta - c_1c_3e^{-c_1x}f\cos\theta \nonumber\\
&& v_{xz} = \frac{e^{c_1x}}{c_3}(v_{xx} + c_1v_x)\cos\theta + c_1c_3e^{-c_1x}f\sin\theta.
 \label{v_ij}
\end{eqnarray}

Another type of integrability conditions arises from the utilization of the relation $u_t=v$. We differentiate with respect to $t$ equations \eqref{u_xx}.
The first equation in \eqref{u_xx} yields
\begin{equation}
  v_{xx} = -c_1v_x + c_3^2e^{-2c_1x}\left(\frac{f_\zeta v_x}{c_1u_x+c_2} + f_t\right)
 \label{Delv}
\end{equation}
while the two last equations give the same results as \eqref{v_ij}. The same result \eqref{Delv} we obtain by differentiating  with respect to $t$ the equation $\Delta[u]=f(\zeta,t)$. By using the result \eqref{Delv} in \eqref{v_ij} we obtain
\begin{eqnarray}
&& v_{xy} = c_3e^{-c_1x}\left[\left(\frac{f_\zeta v_x}{c_1u_x+c_2} + f_t\right)\sin\theta - c_1f\cos\theta\right] \nonumber\\
&& v_{xz} = c_3e^{-c_1x}\left[\left(\frac{f_\zeta v_x}{c_1u_x+c_2} + f_t\right)\cos\theta + c_1f\sin\theta\right].
 \label{vij}
\end{eqnarray}
The two-component real form \eqref{cmareal} of $CMA$ equation becomes
\begin{equation}
  v_t = \frac{e^{2c_1x}}{c_3^2f}[v_x^2 + (c_1u_x+c_2)^2] + \frac{\varepsilon}{f} - (c_1u_x + c_2)
 \label{ma}
\end{equation}
together with its derivative with respect to $x$
\begin{eqnarray}
  && v_{tx} = \frac{f_\zeta}{f}\left[\frac{v_x^2}{c_1u_x+c_2} - (c_1u_x + c_2) - \frac{\varepsilon c_3^2e^{-2c_1x}}{c_1u_x+c_2}\right]\nonumber\\
  &&\mbox{} + 2\frac{f_t}{f}v_x + 3c_1(c_1u_x + c_2) - c_1c_3^2e^{-2c_1x}f.
 \label{v_tx}
\end{eqnarray}
Differentiating \eqref{v_tx} with respect to $x$ we discover the integrability condition $(v_t)_{xx} - (v_{xx})_t = 0$ of \eqref{ma} and \eqref{Delv} in the form
\begin{eqnarray}
  && f_{tt}f - 2f_t^2 - 4c_1^2f^2 + f_{\zeta\zeta}f - 2f_\zeta^2 + 4c_1f_\zeta f \nonumber\\
  &&\mbox{} + \frac{\varepsilon c_3^2e^{-2c_1x}}{(c_1u_x+c_2)^2} (f_{\zeta\zeta}f - 2f_\zeta^2 - c_1f_\zeta f) = 0.
 \label{v_t_xx}
\end{eqnarray}
Since $u_x$ and $x$ in equations for $f$ are allowed only in the combination $\zeta = \ln(c_1u_x+c_2)/c_1 + x + \zeta_0$, which is the independent variable in $f(\zeta,t)$, the equation \eqref{v_t_xx} splits into two equations
\begin{eqnarray}
  && f_{\zeta\zeta}f - 2f_\zeta^2 - c_1f_\zeta f = 0
 \label{1st} \\
  && f_{tt}f - 2f_t^2 - 4c_1^2f^2 + f_{\zeta\zeta}f - 2f_\zeta^2 + 4c_1f_\zeta f = 0.
 \label{2nd}
\end{eqnarray}
Using \eqref{1st} in the equation \eqref{2nd} we simplify the latter equation to the form
\begin{equation}
  f_{tt}f - 2f_t^2 - 4c_1^2f^2 + 5c_1f_\zeta f = 0.
 \label{2_nd}
\end{equation}
Equation \eqref{1st} by the substitution $g=f_\zeta/f$ is reduced to the first-order separable equation $g_\zeta - g(g+c_1) =  0$ with the solution
$g=c_1/(\gamma(t)e^{-c_1\zeta} - 1)$. The corresponding general solution for $f$ is given by the quadrature
\[\ln{f} = \ln{f_0(t)} + c_1\int\frac{d\zeta}{\gamma(t)e^{-c_1\zeta} -  1}\]
or in the explicit form
\begin{equation}
  f = \frac{f_0(t)}{\gamma(t) - e^{c_1\zeta}}.
 \label{f_zt}
\end{equation}
In the original variables this becomes
\begin{equation}
  f = \frac{f_0(t)}{\gamma(t) - c_0e^{c_1x}(c_1u_x+c_2)}.
 \label{ftux}
\end{equation}
Next we substitute the expression \eqref{f_zt} for $f$ in the equation \eqref{2_nd} and after cancelation of the common factor $\gamma - e^{c_1\zeta}$ we obtain the result
\begin{eqnarray}
  && (f_0''f_0 - 2{f'_0}^2 - 4c_1^2f_0^2)(\gamma - e^{c_1\zeta})   \nonumber\\
  &&\mbox{} + 2f_0{f'_0}{\gamma'} - f_0^2\gamma'' + 5c_1^2f_0^2 e^{c_1\zeta} = 0
 \label{2a}
\end{eqnarray}
where primes denote derivatives of functions depending on $t$ only.    Terms with the first and zeroth powers of $e^{c_1\zeta}$ should vanish separately. The terms with $e^{c_1\zeta}$ yield the equation
\begin{equation}
  f''_0f_0 - 2 {f'_0}^2 - 9c_1^2f_0^2 = 0
 \label{pow2}
\end{equation}
with the general solution
\begin{equation}
  f_0 = \frac{1}{\mu(t)},\quad \textrm{where}\quad \mu'' = - 9c_1^2\mu
 \label{f_0}
\end{equation}
or, explicitly,
\begin{equation}
\mu(t) = \mu_1\cos(3c_1t) + \mu_2\sin(3c_1t)
\label{mu}
\end{equation}
with arbitrary real constants $\mu_1$ and $\mu_2$.
The remaining equation consists of the terms in \eqref{2a} without $e^{c_1\zeta}$ which with the use of \eqref{f_0} becomes
\begin{equation}
  \gamma'' + 2\frac{\mu'}{\mu}\gamma' - 5c_1^2\gamma = 0.
 \label{2b}
\end{equation}
For the integration of \eqref{2b} we use its two commuting Lie point symmetries
\begin{equation}
  X_1 = \gamma\partial_{\gamma},\quad X_2 = - \partial_t + \frac{\mu'}{\mu}\gamma\partial_{\gamma}.
 \label{sym}
\end{equation}
Using the appropriate algorithm for the case $G_2Ia$ from H. Stephani's book \cite{steph} we easily obtain the general solution
\begin{equation}
  \gamma(t) = \frac{\nu(t)}{\mu(t)},\quad \textrm{where}\quad \nu'' = - 4c_1^2\nu
 \label{sol2b}
\end{equation}
or, explicitly
\begin{equation}
\nu(t) = \nu_1\cos(2c_1t) + \nu_2\sin(2c_1t)
\label{nu}
\end{equation}
whereas $\mu$ is defined in \eqref{mu} and $\nu_1$, $\nu_2$ are arbitrary real constants.

Using our results in \eqref{f_zt} we obtain the final result for $f$
\begin{equation}
  f(\zeta,t) = \frac{1}{\nu(t) - \mu(t)e^{c_1\zeta}}
 \label{fin f}
\end{equation}
or, alternatively, by using $e^{c_1\zeta_0} = c_0$ in the definition $\zeta = \ln(c_1u_x+c_2)/c_1 + x + \zeta_0$ coming from  \eqref{Delu} and \eqref{om}
\begin{equation}
  f = \frac{1}{\nu(t) - \mu(t)e^{c_1x}c_0(c_1u_x+c_2)}
 \label{ftu_x}
\end{equation}
where $\mu(t)$ is defined by \eqref{f_0} and $\nu(t)$ defined in \eqref{sol2b}. It turns out that the integrability conditions $(v_{xt})_y = (v_{xy})_t$
and $(v_{xt})_z = (v_{xz})_t$ of the equations \eqref{vij} and \eqref{v_tx} with $f$ defined by  \eqref{ftu_x} are identically satisfied without further constraints on the unknowns $v_x$, $u_x$, $\mu$ and $\nu$.

\section{Integration of equations}
\setcounter{equation}{0}
 \label{sec-integr}

After making sure that the integrability conditions of our equations are satisfied we may proceed to the integration of these equations. We start with the integration of the first equation in \eqref{u_xx} with respect to $x$ with the result
\[\frac{1}{2}c_0\mu e^{2c_1x}(c_1u_x + c_2)^2 - \nu e^{c_1x}(c_1u_x + c_2) = c_3^2e^{-c_1x} - \sigma(y,z,t)\]
which can be rewritten as
\begin{equation}
  c_1u_x + c_2 = \frac{1}{c_0\mu}e^{-c_1x}\left(\nu \pm \Gamma^{1/2}\right)
 \label{u_x}
\end{equation}
where
\begin{equation}
  \Gamma = 2c_0\mu\left(c_3^2e^{-c_1x} - \sigma(y,z,t)\right) + \nu^2
 \label{Lamb}
\end{equation}
 with $\sigma(y,z,t)$ playing the role of the constant of integration with respect to $x$. Henceforth we are free to use either upper or lower sign in all the formulas. Using \eqref{u_x} together with the notation \eqref{Lamb} in the formula \eqref{ftu_x} for $f$ we finally obtain
\begin{equation}
 a = \Delta(u) = f = \mp\Gamma^{-1/2}
 \label{f}
\end{equation}
so that \eqref{u_x} becomes
\begin{equation}
  u_x = \frac{1}{c_0c_1}e^{-c_1x}\left(\gamma - \frac{1}{\mu a}\right) - \frac{c_2}{c_1}.
 \label{ux}
\end{equation}

Next we utilize the result \eqref{ux} in the second and third equations in \eqref{u_xx}, using $\Gamma_y=-2c_0\mu\sigma_y$ and $\Gamma_z=-2c_0\nu\sigma_z$ to obtain $\sigma_y=c_1c_3\sin\theta$ and $\sigma_z=c_1c_3\cos\theta$ and hence
\begin{equation}
  \sigma = c_1c_3(y\sin\theta + z\cos\theta) + \sigma_0(t)
 \label{sigm}
\end{equation}
which implies $\sigma_t = c_1^2c_3(z\sin\theta - y\cos\theta) + \sigma'_0(t)$. The result \eqref{sigm} should be used in the definition \eqref{Lamb} of $\Gamma$.

Total derivative of \eqref{ux} with respect to $t$ yields
\begin{equation}
  v_x = \frac{e^{-c_1x}}{c_0c_1}\left(\gamma - \frac{1}{\mu a}\right)_t .
  \label{vx}
\end{equation}
Integrating \eqref{ux} with respect to $x$ we obtain
\begin{equation}
  u = - \frac{\gamma}{c_0c_1^2}e^{-c_1x} + \frac{1}{3c_0^2c_1^2c_3^2\mu^2a^3} - \frac{c_2}{c_1}x + \rho(y,z,t)
 \label{u}
\end{equation}
with the ``constant of integration'' $\rho(y,z,t)$. Differentiation of \eqref{u} with respect to $t$ yields the result
\begin{equation}
 v = u_t = - \frac{\gamma'}{c_0c_1^2}e^{-c_1x} + \frac{1}{3c_0^2c_1^2c_3^2} \left(\frac{1}{\mu^2a^3}\right)_t  + \rho_t.
 \label{v}
\end{equation}
Now we use the expression \eqref{v} to compute $v_y$ and $v_z$ in the first two equations \eqref{v_} which imply the following equations
\[\rho_{ty}(y,z,t) = \frac{1}{c_0c_1c_3}\left(\gamma\sin\theta\right)',\quad
\rho_{tz}(y,z,t) = \frac{1}{c_0c_1c_3}\left(\gamma\cos\theta\right)'\]
with the final result
\begin{equation}
  \rho(y,z,t) = \frac{\gamma}{c_0c_1c_3}(y\sin\theta + z\cos\theta) + \rho_0(t) + r(y,z)
 \label{ro}
\end{equation}
with arbitrary $\rho_0(t)$ and $r(y,z)$. Equation $\Delta[u] = f$ implies $\Delta[r(y,z)] = 0$.

We differentiate $v$ in \eqref{v} with respect to $t$ and use $v_t$ in the third equation of \eqref{v_} which is equivalent to our basic complex Monge-Amp\`ere equation in the two-component form \eqref{2comp} (since $u_t\equiv v$ by construction). This equation after numerous cancelations, an appropriate use of equations \eqref{mu}, \eqref{nu} and \eqref{2b} for $\mu$, $\nu$ and $\gamma$, respectively, and taking two quadratures with respect to $t$ yields the result
\begin{equation}
  \rho_0(t) = \frac{\gamma}{c_0c_1^2c_3^2}\left(\sigma_0 - \frac{\gamma\nu}{3c_0}\right) + \frac{\varepsilon\nu}{4c_1^2} + \rho_1t + \rho_2
 \label{cma2}
\end{equation}
with arbitrary real constants $\rho_1$ and $\rho_2$ and  $\sigma_0(t)$ satisfying the equation
 \begin{equation}
  \sigma_0''(t) + c_1^2\sigma_0(t) = - \varepsilon c_0c_1^2c_3^2\mu.  
 \label{sig0}
\end{equation}
The solution to \eqref{sig0} reads
\begin{equation}
  \sigma_0(t) = \frac{\varepsilon c_0c_3^2}{8}\mu + A\cos{c_1t} + B\sin{c_1t}
\label{si0}
\end{equation}
where $\mu = \mu_1\cos{3c_1t} + \mu_2\sin{3c_1t}$, $A$ and $B$ are arbitrary real constants.

By using $a$, $\sigma$, $\sigma_0$, $\rho$ and $\rho_0$ defined in \eqref{f}, \eqref{sigm}, \eqref{si0}, \eqref{ro}, and \eqref{cma2}, respectively, in \eqref{u} and \eqref{v} for $u$ and $v$ we arrive at the solution \eqref{main}, \eqref{aa} announced at the Introduction.

\section{The metric}
\setcounter{equation}{0}
 \label{sec-metr}

All our equations being completely solved, we can explicitly construct the corresponding metric \eqref{realmetr}.
By using the definition of $Q$ from \eqref{not}, we use the $CMA$ system in the two-component form
\begin{equation}
u_t=v,\quad v_t + u_{xx} = Q \equiv (b^2+c^2+\varepsilon)/a
  \label{CMA2}
\end{equation}
so that for any solution of \eqref{CMA2} the metric \eqref{realmetr} becomes
\begin{eqnarray}
&& {\rm d} s^2 = Q({\rm d} t^2 + {\rm d} x^2) + a({\rm d} y^2 + {\rm d} z^2) \nonumber\\
&&\mbox{} - 2b({\rm d} t\,{\rm d} z  - {\rm d} x\,{\rm d} y) + 2c({\rm d} t\,{\rm d} y  + {\rm d} x\,{\rm d} z).
 \label{metrQ}
\end{eqnarray}
By introducing the coframe of 1-forms
\begin{eqnarray}
&& e^0 = Q^{1/2}\textrm{d}t + Q^{-1/2}(c\textrm{d}y - b\textrm{d}z)\nonumber\\
&& e^1 = Q^{1/2}\textrm{d}x + Q^{-1/2}(b\textrm{d}y + c\textrm{d}z)\nonumber\\
&& e^2 = Q^{-1/2}\textrm{d}y,\quad e^3 = Q^{-1/2}\textrm{d}z.
 \label{e^i}
\end{eqnarray}
we convert the metric \eqref{metrQ} into the diagonal quadratic form
\begin{equation}
  \textrm{d}s^2 = (e^0)^2 + (e^1)^2 + \varepsilon\left[(e^2)^2 + (e^3)^2\right],\quad \varepsilon = \pm 1.
 \label{quadr}
\end{equation}

For our solution, the metric coefficients in \eqref{metrQ} and \eqref{e^i} are defined as follows
\begin{equation}
   a = \Delta(u) = \mp\Gamma^{-1/2}
 \label{aaa}
\end{equation}
\begin{equation}
 - b = v_z - u_{xy} = \frac{1}{c_0c_1c_3}\left[\left(\gamma -\frac{1}{\mu a}\right)\cos\theta\right]_t - c_3e^{-c_1x}a\sin\theta
 \label{-b}
\end{equation}
\begin{equation}
  c = v_y + u_{xz} = \frac{1}{c_0c_1c_3}\left[\left(\gamma -\frac{1}{\mu a}\right)\sin\theta\right]_t + c_3e^{-c_1x}a\cos\theta
 \label{c}
\end{equation}
\begin{eqnarray}
&& Q = \frac{1}{a}\left\{\frac{1}{c_0^2c_1^2c_3^2}\left[\left(\gamma -\frac{1}{\mu a}\right)_t\right]^2\right.
+ \frac{1}{c_0^2c_3^2}\left(\gamma -\frac{1}{\mu a}\right)^2\nonumber\\
&& - \left. \frac{2a}{c_0}e^{-c_1x}\left(\gamma -\frac{1}{\mu a}\right) + c_3^2a^2e^{-2c_1x} + \varepsilon\right\}.
 \label{Q}
\end{eqnarray}
Here $\Gamma$ is defined in \eqref{Lamb}, $\gamma=\nu/\mu$ with $\mu$, $\nu$ and $\sigma$ given in \eqref{mu}, \eqref{nu} and \eqref{sigm}, respectively,  $\theta=-c_1t + \theta_0$.

Since $\sigma$ and $\sigma_t$ involve two different combinations of $y$ and $z$ and there is obviously no reduction in either $x$ or $t$, there is no symmetry reduction of the metric \eqref{realmetr} in the number of independent variables and hence this metric does not admit any Killing vectors.\footnote{An outline of the rigorous proof is given in the Appendix.}

Thus, we have obtained the Ricci-flat (anti-)self-dual metric without Killing vectors. It is generated by a non-invariant solution of the complex Monge-Amp\`ere equation determined solely by its invariance under the pair of special first nonlocal symmetries of $CMA$ without any additional assumptions. We see that such an invariance does not lead to a reduction in the number of independent variables in the solution, as opposed to the invariance under Lie point symmetries. This explains our particular interest for nonlocal symmetry flows in the hierarchies of bi-Hamiltonian systems of Monge-Amp\`ere type which we constructed recently \cite{S_Y}.

\section{Levi-Civita connection 1-forms}
\setcounter{equation}{0}
 \label{sec-connect}

Here we restrict ourselves to the case of the Euclidean signature $\varepsilon = +1$ important for gravitational instantons..

For Levi-Civita connection 1-forms $\omega_{ab}$, Maurer-Cartan structure equations in the absence of torsion imply
\begin{eqnarray}
 && \textrm{d}e^a = c^a_{0i}e^0\wedge e^i + c^a_{23}e^2\wedge e^3 + c^a_{31}e^3\wedge e^1 + c^a_{12}e^1\wedge e^2 = - \omega^a_b\wedge e^b\nonumber\\
 && \omega_{ab} = - \omega_{ba}
 \label{Cartan}
\end{eqnarray}
where $\wedge$ denotes wedge (exterior) product and summation over the dummy index $i$ from $0$ to $3$ is assumed. By solving equations \eqref{Cartan}, one obtains the expressions for the connection 1-forms in terms of the coefficients $c^a_{bj}$\\ (see \cite{egh}, pp. 377, 378)
\begin{eqnarray}
&& \omega^0_1 = - c^0_{01}e^0 - c^1_{01}e^1 + \frac{1}{2}(c^0_{12} - c^1_{02} - c^2_{01})e^2 - \frac{1}{2}(c^0_{31} + c^3_{01} + c^1_{03})e^3\nonumber\\
&& \omega^0_2 = - c^0_{02}e^0 - \frac{1}{2}(c^0_{12} + c^1_{02} + c^2_{01})e^1 - c^2_{02}e^2 + \frac{1}{2}(c^0_{23} - c^2_{03} - c^3_{02})e^3\nonumber\\
&& \omega^0_3 = - c^0_{03}e^0 + \frac{1}{2}(c^0_{31} - c^3_{01} - c^1_{03})e^1 - \frac{1}{2}(c^0_{23} + c^2_{03} + c^3_{02})e^2 - c^3_{03}e^3\nonumber\\
&& \omega^2_3 = \frac{1}{2}(c^3_{02} - c^0_{23} - c^2_{03})e^0 + \frac{1}{2}(c^2_{31} + c^3_{12} - c^1_{23})e^1 - c^2_{23}e^2 - c^3_{23}e^3\nonumber\\
&& \omega^3_1 = \frac{1}{2}(c^1_{03} - c^0_{31} - c^3_{01})e^0 - c^1_{31}e^1 + \frac{1}{2}(c^3_{12} + c^1_{23} - c^2_{31})e^2 - c^3_{31}e^3\nonumber\\
&& \omega^1_2 = \frac{1}{2}(c^2_{01} - c^0_{12} - c^1_{02})e^0 - c^1_{12}e^1 - c^2_{12}e^2 + \frac{1}{2}(c^1_{23} + c^2_{31} - c^3_{12})e^3.\nonumber\\
 \label{omeg}
\end{eqnarray}
Coefficients $c^a_{bj}$ are determined by taking exterior derivatives of $e^a$ in \eqref{e^i}.

To exhibit the results in a compact form, we introduce the notation $\Psi = Q^{-1/2}$ and define
\begin{eqnarray}
&& K = (b\Psi)_t + c\Psi_x + (\Psi^{-1})_z,\quad L = (c\Psi)_t - b\Psi_x - (\Psi^{-1})_y\nonumber\\
&& M = b\Psi_t - c\Psi_x - (\Psi^{-1})_z,\quad N = c\Psi_t + b\Psi_x + (\Psi^{-1})_y.\nonumber\\
\end{eqnarray}
Then the only nonzero coefficients have the form
\begin{eqnarray}
&& c^0_{01} = \Psi_x,\quad c^0_{02} = L,\quad c^0_{03} = - K,\quad c^0_{31} = (b\Psi)_x + b\Psi_x\nonumber\\
&& c^0_{12} = (c\Psi)_x + c\Psi_x,\quad c^0_{23} = - 2\Psi_x,\quad c^1_{01} = - \Psi_t,\quad c^1_{02} = (b\Psi)_t + b\Psi_t\nonumber\\
&& c^1_{03} = (c\Psi)_t + c\Psi_t,\quad c^1_{31} = - K,\quad c^1_{12} = - L,\quad c^1_{23} = 2\Psi_t,\quad c^2_{02} = \Psi_t\nonumber\\
&& c^2_{12} = \Psi_x,\quad c^2_{23} = - M,\quad c^3_{03} = \Psi_t,\quad c^3_{31} = - \Psi_x,\quad c^3_{23} = - N.
 \label{cijk}
\end{eqnarray}
Using the relations inverse to \eqref{e^i}
\begin{eqnarray}
&& dt = \Psi(e^0 - ce^2 + be^3),\quad dx = \Psi(e^1 - be^2 - ce^3)\nonumber\\
&& dy = \Psi^{-1}e^2,\quad dz = \Psi^{-1}e^3
\label{invers}
\end{eqnarray}
we obtain the final form of the solution \eqref{omeg} for connection 1-forms
\begin{eqnarray}
&& \omega^2_3 = \Psi_xe^0 - \Psi_te^1 + Me^2 + Ne^3 = - \omega^0_1\nonumber\\
&& \omega^3_1 = Le^0 + Ke^1 + \Psi_te^2 + \Psi_xe^3 = - \omega^0_2\nonumber\\
&& \omega^1_2 = - Ke^0 + Le^1 - \Psi_xe^2 + \Psi_te^3 = - \omega^0_3.
 \label{oms}
\end{eqnarray}
Due to the relations between $\omega^a_b$ in \eqref{oms}, the equations \eqref{Cartan} imply
\begin{eqnarray}
&& \textrm{d}e^0 = - (\omega^0_1\wedge e^1 + \omega^0_2\wedge e^2 + \omega^0_3\wedge e^3)\nonumber\\
&& \textrm{d}e^1 = \omega^0_1\wedge e^0 + \omega^0_3\wedge e^2 - \omega^0_2\wedge e^3\nonumber\\
&& \textrm{d}e^2 = \omega^0_2\wedge e^0 - \omega^0_3\wedge e^1 + \omega^0_1\wedge e^3\nonumber\\
&& \textrm{d}e^3 = \omega^0_3\wedge e^0 + \omega^0_2\wedge e^1 - \omega^0_1\wedge e^2.
 \label{de^a}
\end{eqnarray}

\section{Riemann curvature}
\setcounter{equation}{0}
 \label{sec-curv}

 Riemann curvature 2-forms are determined by the Cartan equations
\begin{equation}
  R^a_{\ b} = \textrm{d}\omega^a_{\ b} + \omega^a_{\ c}\wedge\omega^c_{\ b}\quad\textrm{where}\quad R^a_{\ b} = \frac{1}{2} R^a_{\ bcd}e^c\wedge e^d
 \label{Rab}
\end{equation}
and $R^a_{\ bcd}$ are the Riemann tensor components.
Due to the relations between connection 1-forms in \eqref{oms}, we obtain from \eqref{Rab}
\begin{eqnarray}
&& R^0_{\ 1} = - R^2_{\ 3} = \textrm{d}\omega^0_{\ 1} + 2\omega^0_{\ 2}\wedge\omega^0_{\ 3}\nonumber\\
&& R^0_{\ 2} = R^1_{\ 3} = \textrm{d}\omega^0_{\ 2} + 2\omega^0_{\ 3}\wedge\omega^0_{\ 1}\nonumber\\
&& R^0_{\ 3} = - R^1_{\ 2} = \textrm{d}\omega^0_{\ 3} + 2\omega^0_{\ 1}\wedge\omega^0_{\ 2}.
 \label{R_ab}
\end{eqnarray}
The symmetry $R^a_{\ b} = - R^b_{\ a}$ implies that the formulas \eqref{R_ab} determine in principle all nonzero curvature 2-forms. Utilizing explicit expressions \eqref{oms} for connection forms we obtain the following expressions for the Riemann tensor components
\begin{eqnarray}
&& R^0_{\ 101} = \Psi_{tt}\Psi - \Psi_t^2 + \Psi_{xx}\Psi - \Psi_x^2 + 2(K^2 + L^2)\nonumber\\
&& R^0_{\ 102} = \Psi^{-1}\Psi_{xy} - \Psi(\Psi_{tx}c + \Psi_{xx}b + M_t) + 3(\Psi_tK - \Psi_xL)\nonumber\\
&& R^0_{\ 103} = \Psi^{-1}\Psi_{xz} + \Psi(\Psi_{tx}b - \Psi_{xx}c - N_t) + 3(\Psi_tL + \Psi_xK)\nonumber\\
&& R^0_{\ 112} = - \Psi^{-1}\Psi_{ty} + \Psi(\Psi_{tx}b + \Psi_{tt}c - M_x) - 3(\Psi_tL + \Psi_xK)\nonumber\\
&& R^0_{\ 113} = - \left[\Psi^{-1}\Psi_{tz} + \Psi(\Psi_{tt}b - \Psi_{tx}c + N_x) - 3(\Psi_tK - \Psi_xL)\right]\nonumber\\
&& R^0_{\ 123} = \Psi^{-1}(M_z - N_y) + \Psi\left[b(M_t + N_x) + c(N_t - M_x)\right] \nonumber\\
&& + M^2 + N^2 + 4(\Psi_t^2 + \Psi_x^2)\nonumber\\
&& R^0_{\ 201} = \Psi (L_x - K_t) + 3(\Psi_tK - \Psi_xL) \nonumber\\
&& R^0_{\ 202} = \Psi^{-1}L_y - \Psi(cL_t + bL_x) - (K^2 + L^2 + 3KM)\nonumber\\
&& - \Psi_{tt}\Psi - \Psi_t^2 + 2\Psi_x^2\nonumber\\
&& R^0_{\ 203} = \Psi^{-1}L_z + \Psi(bL_t - cL_x) - 3(\Psi_t\Psi_x + KN) - \Psi_{tx}\Psi\nonumber\\
&& R^0_{\ 212} = \Psi^{-1}K_y - \Psi(cK_t + bK_x) - 3(\Psi_t\Psi_x - LM) - \Psi_{tx}\Psi\nonumber\\
&& R^0_{\ 213} = \Psi^{-1}K_z + \Psi(bK_t - cK_x) - K^2 - L^2 + 3LN\nonumber\\
&& - \Psi_{xx}\Psi - \Psi_x^2 + 2\Psi_t^2\nonumber\\
&& R^0_{\ 223} = \Psi^{-1}(\Psi_{tz} - \Psi_{xy}) + b\Psi(\Psi_{tt} + \Psi_{xx}) + 2(\Psi_xL - \Psi_tK)\nonumber\\
&& - \Psi_tM - \Psi_xN.
 \label{riemann}\\
&& R^0_{\ 301} = - \Psi(K_x + L_t) + 3(\Psi_xK + \Psi_tL)\nonumber\\
&& R^0_{\ 302} = - \Psi^{-1}K_y + \Psi(cK_t + bK_x) + \Psi\Psi_{tx} + 3(\Psi_t\Psi_x - LM)\nonumber\\
&& R^0_{\ 303} = - \Psi^{-1}K_z - \Psi(bK_t - cK_x) - K^2 - L^2 - 3LN\nonumber\\
&& - \Psi\Psi_{tt} - \Psi_t^2 + 2\Psi_x^2 \nonumber\\
&& R^0_{\ 312} = \Psi^{-1}L_y - \Psi(cL_t + bL_x) + K^2 + L^2 - 3KM\nonumber\\
&& + \Psi\Psi_{xx} + \Psi_x^2 - 2\Psi_t^2 \nonumber\\
&& R^0_{\ 313} = \Psi^{-1}L_z + \Psi(bL_t - cL_x) - \Psi\Psi_{tx} - 3(\Psi_t\Psi_x + KN)\nonumber\\
&& R^0_{\ 323} = - \Psi^{-1}(\Psi_{xz} +\Psi_{ty} ) + c\Psi(\Psi_{tt} + \Psi_{xx})\nonumber\\
&& - 2(\Psi_xK + \Psi_tL) + \Psi_xM - \Psi_tN.\nonumber
\end{eqnarray}
Here $\Psi = Q^{-1/2}$ and the letter subscripts denote partial derivatives.

We note that formulas \eqref{oms} for connection forms and \eqref{riemann} for the Riemann curvature are valid for any hyper-K\"ahler metric
of the form \eqref{metrQ}, so that they can be applied for any solution of the complex Monge-Amp\`ere equation in the two-component form \eqref{2comp}.

For our solution, expressions \eqref{aaa}, \eqref{-b}, \eqref{c}, \eqref{Q} for $a$, $b$, $c$, $Q$ should be used.

\section{Conclusion}

Our search for non-invariant solutions to the elliptic complex Monge-Amp\`ere equation has been motivated  by the fundamental problem of obtaining explicitly the metric of the gravitational instanton  $K3$, since it should not admit any Killing vectors (continuous symmetries). Recently, we produced an example of such a metric, though not an instanton one, by combining our previous approaches to the problem and choosing at random a very particular solution to resulting equations \cite{S-M}.

To the best of our knowledge, there are no more published results on noninvariant solutions of $CMA$ and corresponding metrics of the form \eqref{metrQ} with no Killing vectors.

Here we have demonstrated that a general requirement of invariance under nonlocal symmetries of $CMA$ yields solutions which are not invariant with respect to any point symmetries and therefore no symmetry reduction results in the number of independent variables. We have explicitly constructed such a solution by a meticulous analysis of all integrability conditions of the invariance equations which  made possible a complete integration of these equations with no additional assumptions. Thus, we have obtained the most general form of the solution of $CMA$ which is invariant under the special first nonlocal symmetry in each hierarchy of $CMA$, not just a solution taken out by chance. We have also presented the corresponding ASD Ricci-flat metric without Killing vectors. It has a rather  complicated form and further analysis is needed to study its other properties, such as singularities and asymptotic behavior.

Thus, at this stage we can by no means claim that our metric belongs to $K3$ instanton. However, we have demonstrated how invariance under nonlocal symmetries can be used in the attempts for constructing $K3$.

This method can serve as a direct approach for obtaining noninvariant solutions of $CMA$ from the invariance under other nonlocal flows in the hierarchy. We also point out that all our constructions and results have been obtained simultaneously for elliptic and hyperbolic $CMA$ and so the corresponding metrics have either Euclidean or neutral (ultra-hyperbolic) signature, respectively.

For the Euclidean signature, we have also explicitly calculated the Levi-Civita connection and Riemann curvature. These results are not restricted to  our solution only but more generally are applicable to any other solutions of $CMA$.
Further study of singularities and asymptotic behavior is needed to discover if our solution determines an instanton metric.

\vspace*{2mm}

\section*{Appendix. Point symmetries and noninvariance\\ criterion}
\def\theequation{A.\arabic{equation}}
\setcounter{equation}{0}
 \label{sec-point}

The full set of generators of point symmetries of $CMA$ \eqref{CMA2} has the form
\begin{eqnarray}
&& X_1 = z(t\partial_u + \partial_v),\quad X_2 = (ty - xz)\partial_u + y\partial_v\nonumber\\
&& X_3 = t\partial_t + x\partial_x + u\partial_u,\quad X_4 = y\partial_y + z\partial_z + u\partial_u + v\partial_v\nonumber\\
&& X_5 = \partial_t,\quad X_6 = \partial_x,\quad X_7 = \partial_y,\quad X_8 = \partial_z\nonumber\\
&& X_{\alpha\beta} = (\alpha(t,x) + \beta(y,z))\partial_u + \alpha_t(t,x)\partial_v
 \label{point}
\end{eqnarray}
where $\alpha$ and $\beta$ are arbitrary smooth solutions to the Laplace equations
\begin{equation}
 \alpha_{tt} + \alpha_{xx} = 0,\quad \beta_{yy} + \beta_{zz} = 0.
 \label{laplace}
\end{equation}
The general symmetry generator is a linear combination of particular symmetry generators \eqref{point}
\[X  = \sum\limits_{k=1}^8 c^kX_k + X_{\alpha\beta}\]
with constant coefficients $c^k$. Invariance of the solution to $CMA$
\begin{equation}
  \left(\begin{array}{c}
  u \\ v
  \end{array}\right) = \left(\begin{array}{c}
  F(t,x,y,z) \\ F_t(t,x,y,z)
  \end{array}\right)
 \label{sol}
\end{equation}
with respect to symmetry generator $X$ has the general form
\begin{equation}
 \left. X\left[\left(\begin{array}{c}
  F \\ F_t
  \end{array}\right) - \left(\begin{array}{c}
  u \\ v
  \end{array}\right)\right]\right|_{\begin{array}{c}
  u=F\\
  v=F_t
  \end{array}} =  \left(\begin{array}{c}
  0 \\ 0
  \end{array}\right).
 \label{invcond}
\end{equation}
The explicit form of the equation in the first line of \eqref{invcond} reads
\begin{eqnarray}
&& (c^3t+c^5)F_t + (c^3x+c^6)F_x + (c^4y+c^7)F_y + (c^4z+c^8)F_z \nonumber\\
&& = (c^3+c^4)F + c^1tz + c^2(ty - xz) + \alpha + \beta
 \label{1stline}
\end{eqnarray}
where we use for $F$ the right-hand side of \eqref{main},
whereas the equation in the second line of \eqref{invcond} is the time derivative of \eqref{1stline}.

Applying criterion \eqref{invcond} to our solution with generic values of parameters we find only trivial solution with all $c^k$ and $\alpha,\beta$ to be zero, hence our solution is generically noninvariant which is enough for our purposes here. Nonzero solutions for the constants may happen for special values of our solution parameters which would mean that such special solutions are invariant. The full analysis of the situation is quite lengthy and complicated and deserves a separate study, which is clear from similar analysis of a much simpler solution of the Boyer-Finley equation in \cite{BF}.



\end{document}